\documentclass[onecolumn]{aa}  
\pdfoutput=1
\usepackage{amssymb,amsmath}

\usepackage{natbib}
\usepackage[dvipdfmx]{graphicx}
\usepackage{txfonts}
\usepackage{pdfpages}
\begin{document}

 \title{Polarimetric calibration of large mirrors}

   \author{A. L\'opez Ariste \inst{1}         
          }

   \institute{IRAP - CNRS UMR 5277. 14, Av. E Belin. Toulouse. France\\
              \email{Arturo.LopezAriste@irap.omp.eu}}

   \date{Received ; accepted }

 
  \abstract
   {}
   {To propose a method for the polarimetric calibration of large astronomical mirrors that does not
 require use of special optical devices nor knowledge of the exact polarization properties of the 
 calibration target}
   {We study the symmetries of the Mueller matrix of mirrors to exploit them for polarimetric calibration under the assumptions that only the orientation of the linear polarization plane of the 
   calibration target is known with certainty.}
   {A method is proposed to calibrate the polarization effects of  single astronomical mirrors by the observation of calibration targets with known orientation of the linear polarization. We study the
   uncertainties of the method and the signal-to-noise ratios required for an acceptable calibration. We list astronomical targets ready for the method. We finally extend the method to the calibration of two or more mirrors, in particular to the case when they share the same incidence plane.}
   {}


   \maketitle
%

\section{Introduction}

Polarization is not directly measurable by the common astronomical detectors in the optical part of the spectrum. Its measurement relies on the comparison of
two intensity measurements that have been modulated by the polarization. For this idea to work both measurements have to be identical except for the polarization of the
astronomical object.  This general rule, an axiom of polarimetry one could say, is broken by many sources of spurious polarization and crosstalks among 
different states of polarization. The listing, analysis and correction of all these sources of polarimetric errors constitutes the bottom line of good astronomical
polarimetry \citep{bagnulo_stellar_2009}.

An annoying source of both instrumental polarization and crosstalk is the reflection upon the mirrors of telescopes
\citep{miller_effects_1963,de_juan_ovelar_instrumental_2014}. The difficulty with them is that the concerned mirrors are often the
main  mirrors of the telescope, hence the largest. And this, quite commonly, puts them off-limits for any manipulation. Reflection 
upon mirrors polarizes light as long as there is no spherical symmetry in
the ray path. Prime or Cassegrain foci which preserve the spherical symmmetry of the ray path are mostly or almost completely free of any spurious polarization effect. 
Succesful polarimeters have been put in those places, both in  solar \citep{arnaud_solar_1998} and night telescopes \citep{auriere_stellar_2003,snik_harps_2010,donati_espadons:_2003}.
But for many different reasons, placing a polarimeter at such a favorable place is not always possible. 

When forced to measure polarization behind polarizing optics, the only way out is calibration of those optics. The straightforward manner to calibrate
the polarimetric  effects of telescope optics is to introduce a well known polarization and see how it transforms after passage through the optics. 
Thus in solar telescopes it is a common practice to place in the entrance window polarizers and retarders, and to create a model of the telescope Mueller matrix with a few free 
parameters to be fit \citep{beck_polarization_2005,skumanich_calibration_1997,balasubramaniam_measurement_1985}. Night telescopes can seldom afford 
such calibrations. Most of the night telescopes in use have just too large apertures to be 
covered with homogeneous enough polarizers and or retarders. The awkward solution in such cases is to rely on calibration stars whose polarization 
has been  determined elsewhere and assumed to be constant \footnote{See for example the instrument description page of 
FORS at http://www.eso.org/sci/facilities/paranal/instruments/fors/inst/pola.html}. As both solar and night telescopes grow larger and larger, this is the
situation into which polarimetric calibration is forced to fall.

The use of a calibration target is however undesirable in the sense that it implies a complete confidence in the polarisation measured by 
somebody else with other, perhaps 
non-existent, polarimeters and telescopes \citep{leroy_linear_1995}, or on the physical assumptions about the origin of such polarization. The starting point of the present work is 
to present a method that, in many cases, 
lifts that concern. Indeed it is uncommon that we can determine with enough accuracy the degree of polarization of an astronomical object. However,
we  can often know with exquisite 
accuracy  the orientation of the linear polarization of many objects: Scattering polarization off the lunar surface 
is strictly perpendicular to the plane Sun-Moon-Earth for most of the lunar cycle, and strictly parallel during a few days around full moon. 
The amount of polarization,
however, depends on the point of the surface of the Moon we are looking at, the phase of our satellite, the wavelength and, at any rate, with a not 
too good accuracy.  We cannot trust on the
degree of polarization, but we are certain of the polarization angle of 
this linear polarization source. Can we effectively make use of this information to calibrate  
large telescope mirrors?  We present in Sect. 2 a method to do so for single mirrors. In Sect. 3 we make a list of astronomical targets with the characteristics of 
known orientation of its linear polarization, useful for our method. Finally in Sect. 4 we extend our method, whenever possible, to 
the calibration of more than one mirror, in particular when those mirrors share the same incidence angle.

We anticipate that this method should be of interest for tilted primary mirrors as the one on DKIST (Daniel K. Inouye Solar Telescope,
formerly known as the Advanced Technology Solar Telescope, ATST) , and for Nasmyth focal instruments (ESO's VLT and E-ELT for example). It 
may also be useful also for other non-accessible astronomical instruments as those space-borned.
   

\section{Description of the method}

The proposed calibration method takes advantage of the knowledge of the symmetries of the Mueller matrix of the mirror to be calibrated and the 
existence of a source of linear polarization whose amplitude we do not know but whose orientation is perfectly well determined respect to 
celestial reference frames. We further assume that by rotating the mirror we can vary this orientation respect to the mirror at will.

The Mueller matrix of a mirror is determined by two parameters: $\chi$, the
ratio of the Fresnel transmission coefficients for the parallel and perpendicular electric fields of the ray, and $\Delta$ the phase retardance between those two field components after reflection. $\chi$ and $\Delta$ both 
depend on wavelength and incidence angle, so that the calibration may have to be repeated for different wavelengths and incidence angles, if required. 
In the particular examples, we have considered that a wavelength calibration is still required, but that the incidence angle is fixed by the geometry 
of the optical path of the telescope.

$\chi$ and $\Delta$ can be computed from the knowledge of the complex refraction index of the mirror \citep{capitani_polarization_1989}, but if this
information were available we would not need to calibrate the mirror. More commonly the refraction index of a mirror is not known with sufficient 
precision. Even for metallic mirrors this is not a surprising fact: Aluminum oxidizes over time and layers of dust and dirt may 
also modify the effective refraction index of the mirror; silver is usually covered with protective and enhancing multilayers with not very well determined refraction indices.

We will consider that $\chi (\lambda)$ and $\Delta (\lambda)$ are the parameters to be determined for the complete calibration of the mirror. The
corresponding Mueller matrix at a fixed incidence angle is given by
\begin{equation}
 M_{mirror}=\frac{1}{2}\left( \begin{matrix}  1+\chi^2 & 1-\chi^2 & 0 & 0 \\ 1-\chi^2 & 1+\chi^2 & 0 & 0 \\ 0 & 0 & 2\chi\cos \Delta & 2\chi\sin \Delta  
\\ 0 & 0 & -2\chi\sin \Delta& 2\chi\cos \Delta \end{matrix} \right)
\label{Mmirror}
\end{equation}
where the Stokes parameters have been defined such that Stokes $Q$ is positive when the linear polarization (the direction of oscillation of the electric field) 
is perpendicular to the plane of incidence of the mirror \footnote{See Sections 1.4.2 and 1.5.3 from \cite{born_principles_1999} for a discussion on the conventions of 
plane of polarization.}. We have chosen a particular realisation of this Mueller matrix that may differ from other sources. For example \cite{capitani_polarization_1989} give
$\chi^2 -1$ as the $M_{12}$ term, because they define positive $Q$ to be in the plane of incidence, rather than perpendicular to it. They however coincide with us in the definition 
of $\Delta$. \cite{joos_reduction_2008}, on the other hand, use the same definition of $Q$ and $\chi^2$ while their definition of $\Delta$ is $180^\circ$ away from ours. All 
these conventions are correct as long as one sticks to the adopted definitions, and our method is not dependent on which matrix is used. Our choice in Eq.\ref{Mmirror} has
the advantage of being easily related to a Lorentz transformation, an aspect that has no influence on the technique or the conclusions of this paper, but which is coherent
with the quantum nature of light polarization.

Coming back to the polarimetric calibration of the mirror, the  key points  of this matrix are that, as we said, it is completely 
determined by the two parameters $\chi$ and $\Delta$ and that it is a block-diagonal matrix decoupling the dichroic $2\times 2$ sub-matrix involving
$I$ and $Q$ from the retardance sub-matrix involving $U$ and $V$. This uncoupling of the two effects induced by the mirror in the polarization of light is a key point 
of our proposition.

The second key point is the availability of astronomical sources emitting
linearly polarized light whose amplitude may be unknown, but whose orientation is perfectly fixed out of pure geometrical arguments. In the coming sections we will
list and study some of these sources in detail, but 
to fix ideas let us give at this point two examples. Both cases involve scattering that polarizes light in the plane perpendicular to the scattering plane. If the
incoming light is unpolarized, this is the only source of polarization and hence its orientation is perfectly known from the knowledge of the scattering plane. This is
the case of the continuum
light reflected off the Moon or Mercury \citep{leroy_polarization_2000}: the scattering plane is the plane 
Sun-Moon-Earth and, since the integrated light of the Sun has zero or negligible linear polarization, the reflected light will be linearly polarized perpendicular to this plane.
The degree of polarization depends on the
physical and geometrical properties of the regolith in the Moon and on
the scattering angle. Although there is ample information on this degree of polarization  for different places over the lunar surface, it is often
not precise enough for calibration purposes, leaving aside the difficulty
of pointing at particular lunar spots. We will not make use of that degree of polarization, just the knowledge of the orientation.  Our second 
example can be found in
the solar limb where many lines show linear polarization after atomic resonance scattering in the upper photosphere or low chromosphere 
\citep{stenflo_second_1997,gandorfer_second_2000}. The continuum is also
slightly polarized, but its amplitude is too weak to be of interest \citep{leroy_new_1972}. Some atomic lines however have clear signatures of polarization
of up to  0.1\% the intensity   with an
orientation always parallel to the solar limb. The degree of polarization is small but it is at a comfortable level for present day solar polarimeters.

Defining as positive Stokes $Q$ the direction perpendicular to the scattering plane, in analogy with positive $Q$ being perpendicular to the plane of 
incidence of a mirror, 
the light coming from those objects (the Moon, Mercury, or particular solar lines
observed near the limb) is described by the Stokes vector
\begin{equation}
\vec{J}=\left( \begin{matrix} I \\ Q \\ 0 \\ 0\end{matrix}\right)
 \end{equation}
where $I$ and $Q$ are unknown but constant during the calibration. This assumption of constancy is not particular to our method. In order to see the effects of our mirror
on this light we have to rotate the Stokes vector so that the definitions 
of positive Stokes $Q$ coincide. If we define $\alpha$ as twice the known angle between
the scattering plane and the incidence plane of the mirror\footnote{By defining $\alpha$ as twice the angle we avoid carrying the factor 2 all over
the equations of the paper.}, the 
incoming Stokes vector in the reference frame of the mirror will be 
\begin{equation}
\vec{J}_{in}=\left( \begin{matrix} I \\ Q \cos \alpha \\ Q\sin \alpha \\ 0\end{matrix}\right) 
\end{equation}
By changing $\alpha$ we can decide to illuminate the mirror with linearly polarized light of any orientation and select which part of the Mueller matrix is at work: dichroism,
retardance or any combination of both. This 
is the bottom line of our method. And this is also the critical point: the method requires that we can modify $\alpha$ at will. In many telescopes there is the choice
of the parallactic angle under which a target is observed. We will assume that this is possible and that the telescope rotation involves the mirror which is being
calibrated.  For solar telescopes this is not needed: the polarization plane of lines with atomic
resonance scattering polarization is perpendicular to 
the local limb and thus it rotates as we observe points at all position angles around the solar circumference.

Provided we can measure the reflected light off out mirror at all values of
the angle $\alpha$, we can easily see that the reflected Stokes vector will be
\begin{equation}
\vec{J}_{out}=\frac{1}{2}\left( \begin{matrix} (1+\chi^2)I+(1-\chi^2)Q\cos\alpha \\ (1-\chi^2)I+(1+\chi^2)Q\cos\alpha
\\2 \chi\cos \Delta Q\sin \alpha \\  2 \chi\sin \Delta Q\sin \alpha \end{matrix}\right)
\end{equation}
This light is to be analyzed with a polarimeter sitting behind the mirror and, at this point, before any other uncalibrated optics. From this polarimeter we 
require to provide the following measurements: $(I\pm Q)_{out}$ and $(I \pm U)_{out}$. These are some of the usual measurements
provided by polarimeters. Often the polarization analysis is made by a polarizing beamsplitter as a Savart plate, so that the detector receives
two beams with intensities $I\pm S$,where $S$ is any of the Stokes parameters $Q$,$U$ or $V$, or a combination of them. So that we need two measures times
two beams to complete the required dataset. 
Almost always the polarimeter allows for temporal modulation, so that the four measurements are taken in sequence 
by adequately rotating retarder plates in the polarimeter. And in the most
performant polarimeters all this is done simultaneously to diminish systematic errors. Summarizing, the details on how those 4 measurements are
taken will depend on each particular instrument and can vary to optimize efficiencies \citep{del_toro_iniesta_optimum_2000}.

Explicitly we will have as measurements
\begin{eqnarray}
(I+ Q)_{out}&=& I+Q\cos \alpha  \\
(I- Q)_{out}&=&\chi^2I - \chi^2 Q\cos \alpha \\
(I+ U)_{out}&=&(1+\chi^2)I+(1-\chi^2)Q\cos \alpha +2\chi\cos\Delta Q\sin \alpha \\
(I- U)_{out}&=&(1+\chi^2)I+(1-\chi^2)Q\cos \alpha -2\chi\cos\Delta Q\sin \alpha 
\end{eqnarray}
From this point on it is tempting to introduce a model and let the computer
fit those signals. But let us continue the analytical treatment to verify 
that the calibration does not depend on the actual degree of polarization $Q$. 
Actually at this point we are going to impose a further requirement and it is that we do not depend on $I$ or the actual intensity measurement in our detectors. We
impose these restrictions because there is no reason to pretend that we know the exact intensity coming from our celestial object 
and we cannot let the calibration depend on the accurate photometric calibration of our instruments. Polarization measurements usually reach 
signal-to-noise levels of $10^5$, but absolute photometry and flat-field
calibrations are seldom good beyond $10^3$ levels. Thus we seek our calibration to be insensitive to the absolute photometric calibration.

\begin{figure}
   \centering
   \includegraphics[width=\hsize]{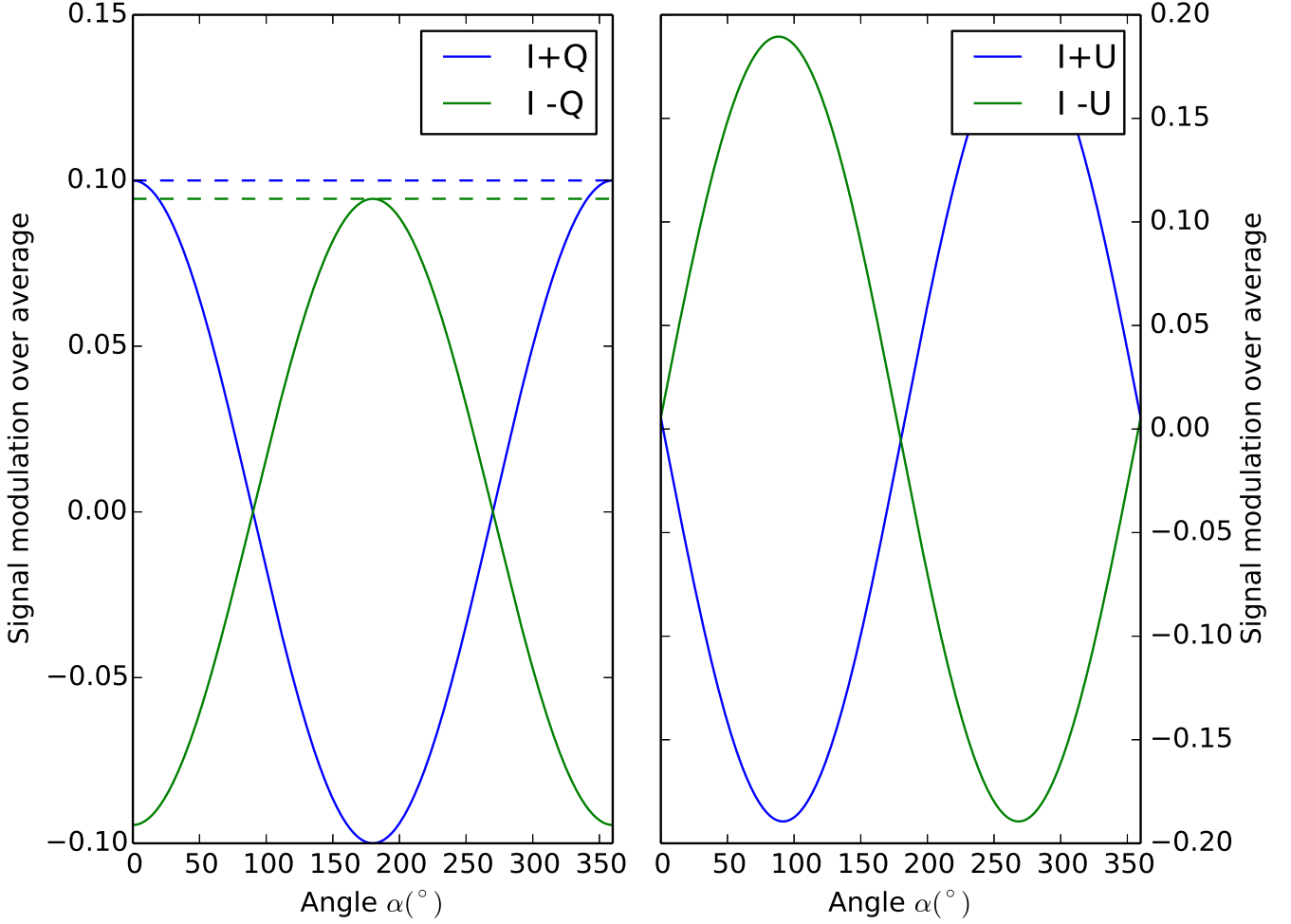}
      \caption{Examples of measures $I\pm Q$(left) and $I\pm U$(right) for
      the case of an Aluminum mirror at $45^\circ$ incidence of light with a wavelength of 5000\AA~ and a degree of polarization of 10\%.
      In all cases the average intensity has been subtracted.
              }
         \label{QU}
   \end{figure}

As we measure for different values of $\alpha$ we see that $(I+ Q)_{out}$ and $(I - Q)_{out}$ (see Fig.\ref{QU}) vary sinusoidally around 
constant levels, $I$ and $\chi^2 I$, which depend on the intensity $I$ and the photometric calibration. This constant level can
however be easily subtracted from both measurements independently by averaging over all the measurements.
And, once this average is subtracted,
the amplitude of the variation, the contrast, is independent of the photometric calibration,  since given by the
unknown $Q$ in one
case and the product $Q\chi^2$ in the other. The ratio of the two contrasts
gives us directly $\chi^2$ one of our mirror parameters, independently of $Q$.

The second pair of measurements around $U_{out}$ starts by a similar description. Both are given by sinusoidal variations with 
$\alpha$ around a constant value $(1+\chi^2)I$, that we subtract before continuing. The two curves are given by
\begin{equation}
(1-\chi^2)Q\cos \alpha \pm 2\chi\cos\Delta Q\sin \alpha
\end{equation}
By defining  $A=(1-\chi^2)Q$ and $B=2\chi\cos\Delta Q $
we see that we can rewrite those measurements as
\begin{equation}
\sqrt{A^2+B^2}\cos (\alpha \mp \delta)
\end{equation}
where 
\begin{equation}
\cos \delta = \frac{A}{\sqrt{A^2+B^2}}. 
\end{equation}
The sinusoids of these measurements in Fig.\ref{QU} both have an amplitude of 
$\sqrt{A^2+B^2}=Q\sqrt{(1-\chi^2)^2+4\chi^2 \cos^2\Delta}$
which depends on Q; and a phase shift $\delta$ given by
\begin{equation}
\cos \delta=\frac{1-\chi^2}{\sqrt{(1-\chi^2)^2+4\chi^2 \cos^2\Delta}}
\end{equation}
which does not depend on the value of $Q$ and from which $\Delta$ can be determined provided $\chi^2$ is known. This phase shift has different sign
in one and other measurements $(I\pm U)_{out}$ providing extra redundancy.

The combination of the two measurements, the ratio of contrasts in one case, the phase shift in the other allows us to determine $\chi$ and $\Delta$, and hence to
calibrate the Mueller matrix of the mirror. Both measurements are independent of the degree of polarization $Q$ or the intensity
$I$, or the actual photometric calibration provided it is stable during the process of measurement over sufficient different values of $\alpha$

\section{Required accuracy}

In the previous section we delivered the good news: a calibration method that does not require
any modification of the telescope, just the observation of a celestial object; and which only requires knowledge of the orientation of the linear polarization, but not
the degree of polarization.

Now it is time for the bad news. Although the proposed measurements have the advantage of not depending on the amplitude of polarization, the actual achievable
signal-to-noise ratios do depend on it. 

\begin{figure}
   \centering
   \includegraphics[width=\hsize]{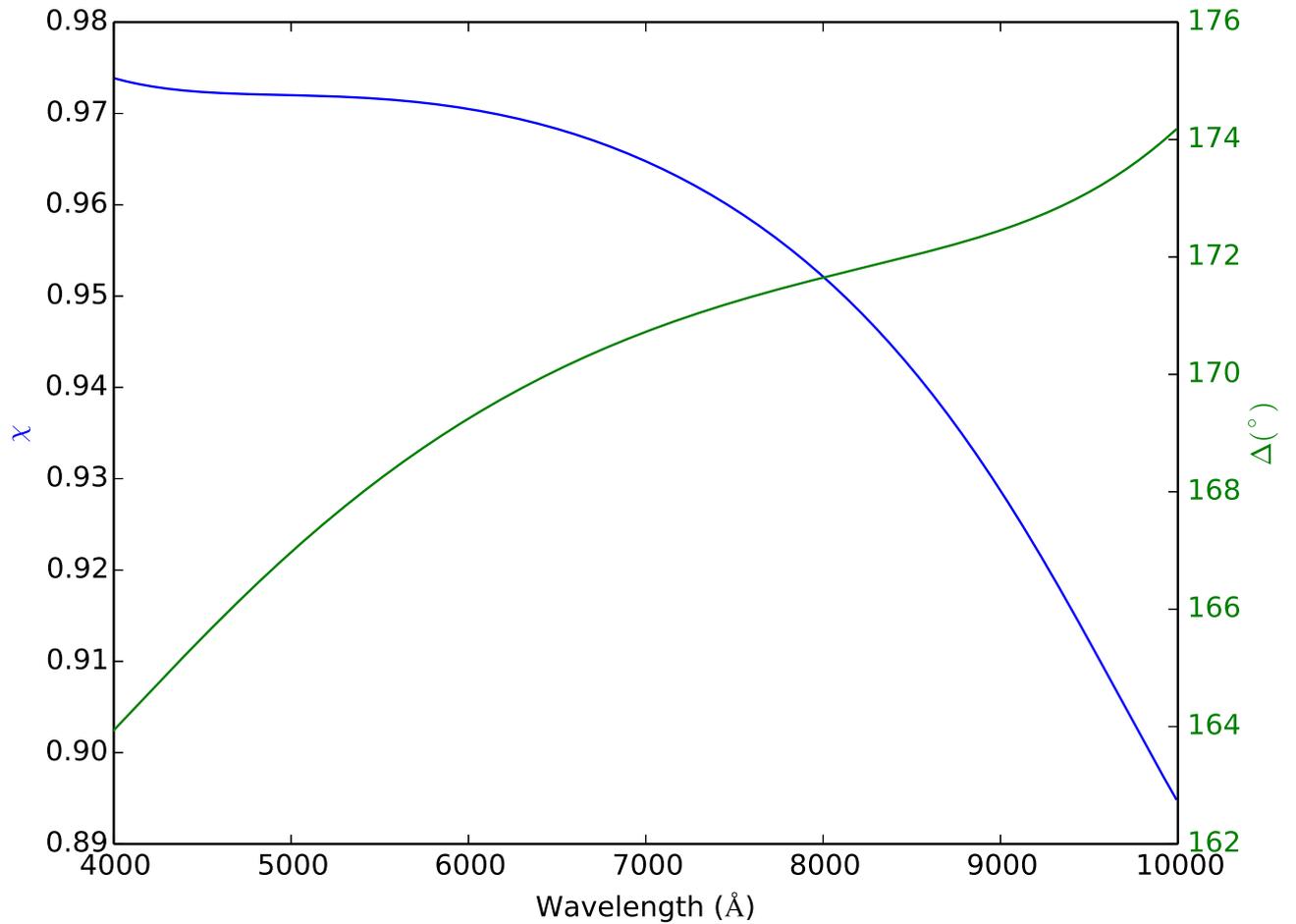}
      \caption{Values of $\chi$ (blue curve, left axis) and $\Delta$ (green curve, right axis) as a function of wavelength for an Aluminum mirror at $45^\circ$ incidence.}
         \label{XDelta}
   \end{figure}

\begin{figure}
   \centering
   \includegraphics[width=\hsize]{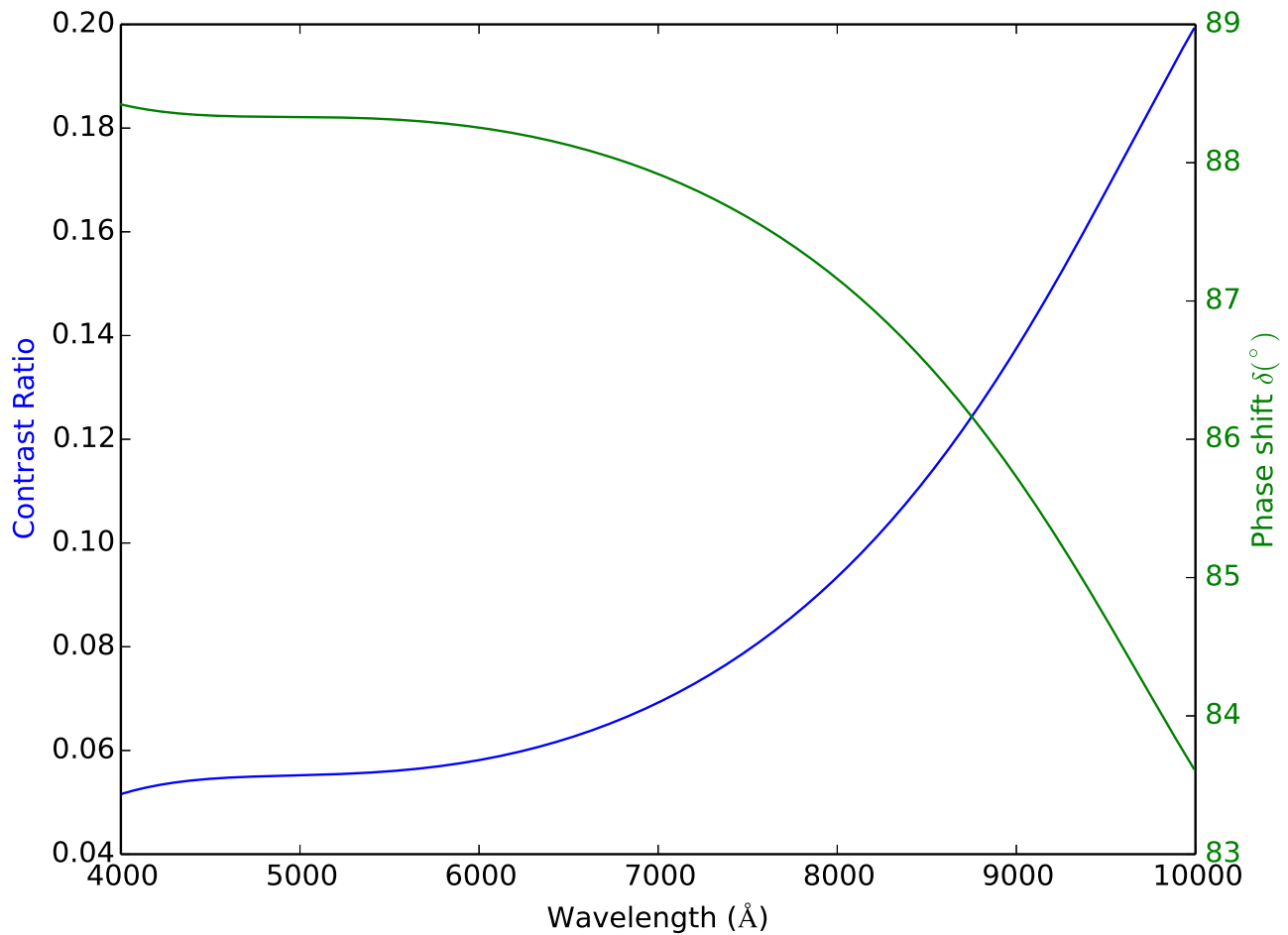}
      \caption{Expected differences in amplitude (blue curve, left axis) and phase (green curve, right axis) between the sinusoids of $I\pm Q$ (left plot of Fig. 1) and
      $I \pm U$ (right plot of Fig. 1) respectively. They have been computed as in Fig. 2 as a function of wavelength for an Aluminum mirror at $45^\circ$ incidence.}
         \label{contrast}
   \end{figure}

Thus, in the case of the measurement of $\chi$ as the ratio of contrasts we notice that the two sinusoids in the left plot of Fig. \ref{QU} have
an amplitude of roughly $Q$, the degree of polarization of the sources and a difference of amplitude of $1-\chi^2$. Fig.\ref{XDelta} shows the typical values of $\chi$ for an
Aluminum mirror at different wavelengths assuming an incidence angle of $45^\circ$. 
They are very close to 1 and hence the differnce in amplitude $1-\chi^2$ between the two sinusoids of Fig. \ref{QU} is really small. Fig.\ref{contrast} shows those
differences in the same conditions of Fig.\ref{XDelta}

\begin{figure}
   \centering
   \includegraphics[width=\hsize]{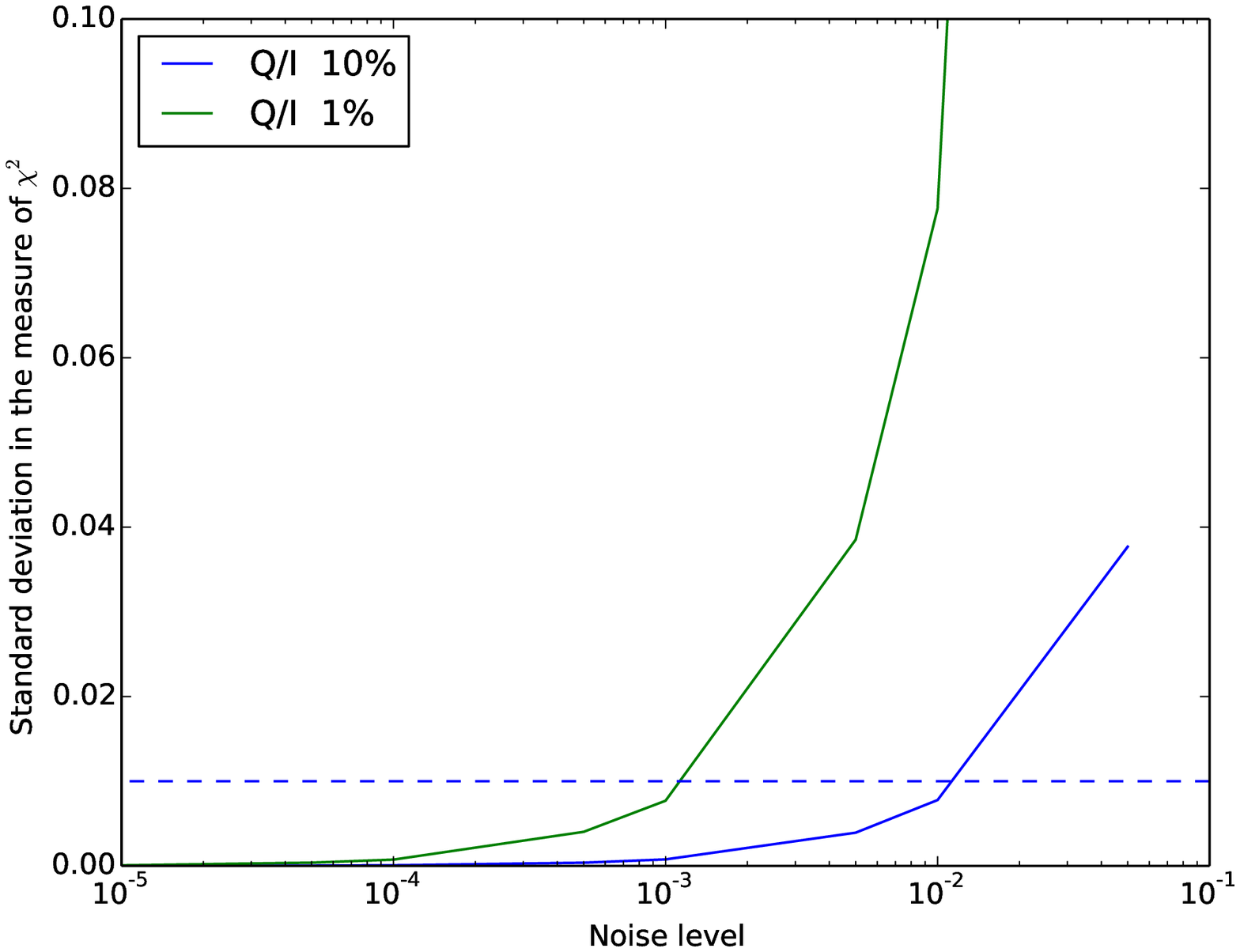}
      \caption{Residual instrumental polarization at 500nm from an Al mirror at $45^{\circ}$ incidence for different Gaussian noise levels.}
         \label{Xnoise}
   \end{figure}

For typical degrees of polarization of the lunar or Mercury's surface of 10\%, the difference in amplitude between the sinusoids to be measured is of $5\times 10^{-3}$ the 
total light, while for the resonance scattering polarization of the \ion{Sr}{I} line over the solar limb, polarized at 1\%, we drop to amplitude differences of $5\times 10^{-4}$.
Measuring those signals implies that noise levels are even lower than that. Fig.\ref{Xnoise} shows an illustration of this issue with the 
calibration of an Aluminum 
mirror at $45^{\circ}$ incidence at 500nm. In a Montecarlo test, Gaussian random noise of the amplitudes shown in the figure is added to the 
signals then they are fit with a Levenberg-Marquardt inversion algorithm.   The quantity plotted is the average  residual  of $\chi^2$ 
along the Montecarlo series after
the measure of the ratio of contrasts. This residual will induce dichroism, also called instrumental polarization, by crosstalk of the intensity into the
linear polarization. The dashed horizontal line gives the requirement for the
maximum instrumental polarization for the DKIST project \footnote{See \textit{SPEC-0009: ATST System Error Budgets document} at dkist.nso.edu} for reference of the
typical accuracies expected in the calibration of the mirror. As a thumb-rule
noise levels should be one order of magnitude below the average linear polarization to respect the typical requirements for the maximum instrumental polarization.

\begin{figure}
   \centering
   \includegraphics[width=\hsize]{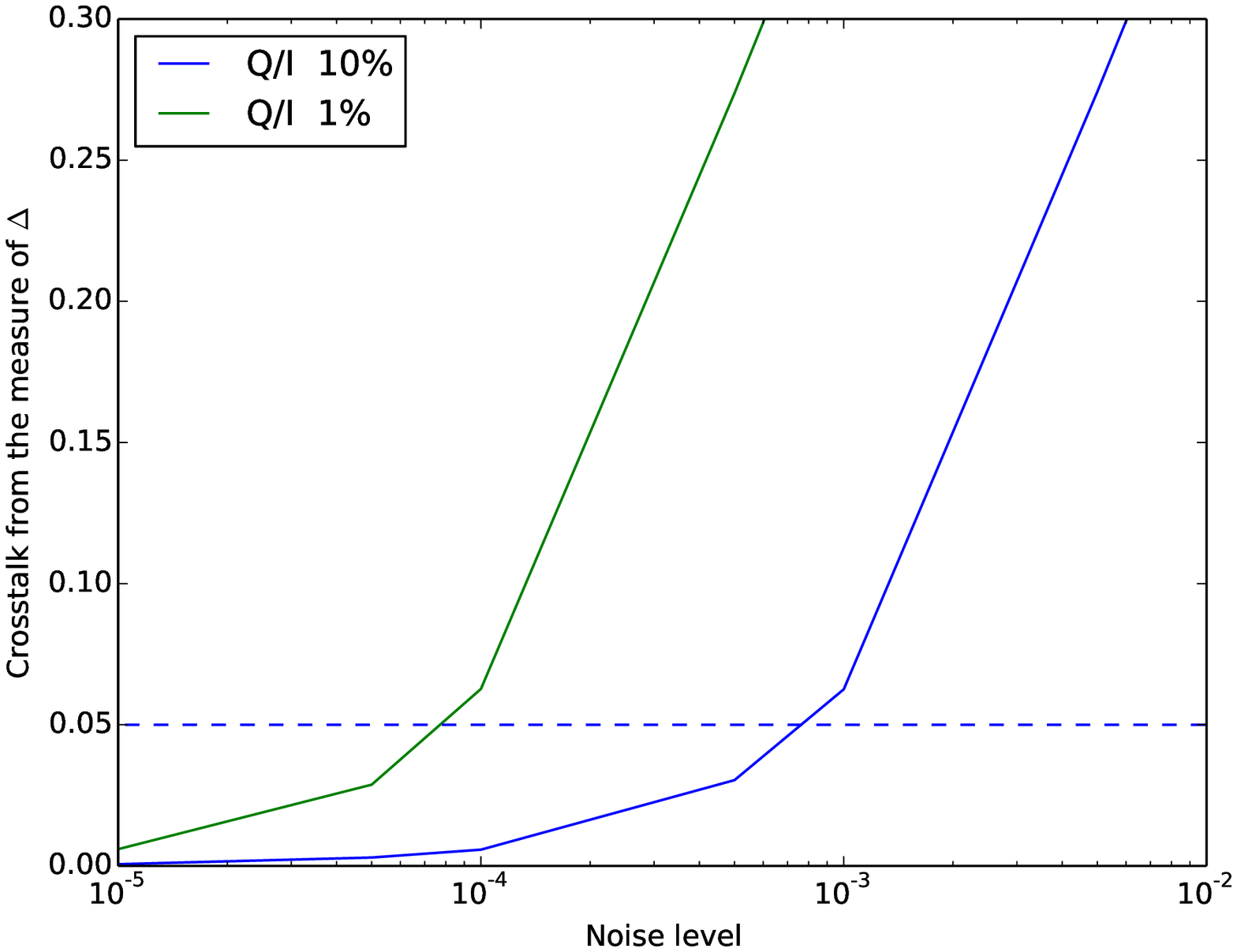}
      \caption{Residual crosstalk at 500nm from an Al mirror at $45^{\circ}$ incidence for different Gaussian noise levels.}
         \label{Deltanoise}
   \end{figure}

Although less evident to explain with a simple arithmetic argument as the one used for the measure of $\chi$, the measure of the retardance $\Delta$ suffers more from
the presence of noise. Fig. \ref{Deltanoise} shows the
equivalent calculation of Fig.\ref{Xnoise} for the retardance. What is 
actually ploted is $2\chi\sin \sigma_{\Delta}$, that is the residual of the
polarization crosstalk in the Mueller matrix for the mirror after measurement
of $\Delta$ from the phase of the sinusoids in Fig.\ref{QU}. The DKIST projet
requirements  are that this crosstalk should not exceed 5\%, a level indicated
with the dashed line. Despite this weaker requirement, compared to the instrumental polarization, the noise levels should be even lower to correctly
determine the retardance of the mirror with this method.

The examples given above show that the absence of a fully polarized source implies the necessity of very high signal-to-noise ratios for the calibration. The noise levels are not beyond reach of present instruments. Most solar spectropolarimeters reach those levels per spectral bin. And yet 
they are low enough that require special attention into how the measurement
is done and analyzed. This is the main trade-off for our proposed calibration
technique: no modification is required, but the measurement has to be very
sensitive.

\section{Celestial targets with known linear polarization orientation}
This section is just a list of possible targets for day and night observation
emitting linearly polarized light in a plane which is exclusively determined
by geometry.

\begin{description}
\item The Moon

It is perhaps the most obvious and available target for both day and night observations, though it may be too brilliant for the largest night telescopes in their
usual observing modes.
Solar light scattered off the lunar regolith is polarized strictly perpendicular to the plane Sun-Moon-Earth and its integrated polarization peaks at almost
20\% polarization in the blue wavelengths at $110^\circ$ phase (or about one week after the waning quarter. The different materials 
over the surface of the moon and its heterogeneous distribution makes this polarization not symmetric around the full moon. Thus, the polarization peaks at a 
mere 15\% during the crescent phase in the blue \citep{dollfus_lunar_1999,dollfus_polarisation_1956,leroy_polarization_2000}.

Because of its large aparent diameter it may also be of interest to point at
particular regions with even larger polarization levels. The albedo-polarization relation suggests that the dark mare are more polarized than 
the bright regions and  will result in smaller errors of calibration.

\item Mercury, Mars and asteroids.

Second to the Moon, the atmosphereless planets are equally interesting targets. Dimmer than the moon, they may be of more interests for large night telescopes, and
yet they are bright enough to be observed in daylight with solar telescopes. Mercury peaks at around 10\% around quadrature. The fast period of its orbit also means
that it is available with maximum polarization roughly once a month, although perhaps too near to the Sun for non-solar telescopes
\citep{dollfus_optical_1974,leroy_polarization_2000}

Mars is a less interesting but also less dangerous object for night telescope. Due to its orbit external to that of the Earth, the maximum 
scattering phase is just $45^\circ$ and this only twice a year at most. This
low scattering phase means that the polarization is low to start with and it
is further reduced by Mie and Rayleigh scattering in the tenuous atmosphere of Mars. Its peak polarisation is just of 1\% or 2\% depending on wavelength 
and region \citep{dollfus_planet_1969,leroy_polarization_2000}.

The same problems of low polarization levels affect minor planets, whose far orbits translate in low scattering angles and weak polarizations. Often 
also asteroids are seen in near $0^\circ$ phase, a region in which, due to
double scattering, the polarization changes sign and is found to be parallel to the plane Sun-asteroid-Earth, rather than perpendicular. On the other hand
they are available at almost any visual magnitude providing the observer 
with a long list of objects available \citep{leroy_polarization_2000,bagnulo_linear_2015,belskaya_polarimetry_2009,penttila_statistical_2005}.

\item The second solar spectrum

For solar telescopes, the second solar spectrum is a nice source of polarization for calibration purposes available directly over the solar surface.  By this
name \citep{gandorfer_second_2000,stenflo_second_1997} called the spectrum of linear polarization of solar lines at or near to the solar limb. The intricacies of 
the atomic resonance scattering polarization translate in that there is almost no correlation between the signals seen in the spectrum of linear polarization near the solar 
limb and the usual intensity spectrum, with strong lines in one absent in the other  and viceversa.

The two main advantages for solar telescopes are its availability (just point 
to the solar limb) and the fact that to rerpoduce the rotation of the angle of polarization $\alpha$ it is sufficient to point at the limb at different position angles, 
since the linear polarization of the second solar spectrum
is parallel (in some special cases perpendicular) to local solar limb. Its main and serious disadvantages are the low amplitude of the signal. The \ion{Sr}{I} line at 
4607\AA~  presents the larger polarization amplitude in the visible spectrum \citep{faurobert_investigation_2001,gandorfer_second_2000} and it is just 1\%. And
contrary to the Moon or planets it is not the continuum that is polarized, but just a line with a relatively small 150m\AA~ equivalent width. Most of 
the lines have signals in the range of $1-5\times 10^{-4}$ all over the spectrum. And this may imply calibrations with very high signal-to-noise ratios.

Detailed analysis of the mechanisms of polarization of these lines show that
at high spatial resolution there is the possibility of appearance of other polarizations. The Hanle effect may rotate the plane of polarization as a function of the
magnetic field and depending on the conditions on the photosphere; and circular polarization may appear if sensible magnetic fields are present around the
observation region. To avoid this the solution is to
degrade spatial and temporal resolutions: this cancels out all those effects and has the extra advantage of increasing the photon flux.

\item The blue sky

We end up this list with a bad source of polarization which however comes easily to mind: the polarization of the daylight sky due to Mie and Rayleigh 
scattering of the solar light. The degree of polarization can be beyond 90\% at $90^\circ$ from the Sun, and its orientation is, as in the previous
cases of scattering polarization, perpendicular to the direction to the Sun of the observed point. However this direction is only approximative, as
it depends on the presence of other sources of spurious illumination as clouds or water surfaces, that can alter it 
\citep{berry_polarization_2004,hegedus_anomalous_2007,horvath_first_2002,lee_digital_1998}. We find that this extra source of certainty in the
orientation of the linear polarization impacts negatively the usefulness of the proposed calibration method.

\end{description}

\section{Two and more mirrors}
In general, to calibrate more than one mirror brings in too many unknowns to the equations. In the usual calibration techniques it is assumed
that all the mirrors are identical. In spite of this simplification the calibration still requires measurements in many of the different possible
configurations of the positions of those mirrors and introducing more than just linear polarization to make the solution non ambiguous and 
robust \citep{skumanich_calibration_1997}. A couple of particular cases however can still be handled in a similar manner to the one proposed in
this work. First, we should mention the case of two crossed mirrors   
studied before \citep{martinez_pillet_proposal_1991,sanchez_almeida_proposal_1995}. The combined Mueller matrix of the two mirrors becomes diagonal
and there is no instrumental polarization. This diagonal matrix introduces differential absorptions into the different Stokes parameters from
which the values of $\chi$ and $\Delta$ can be measured and be used for the computation of the Mueller matrix in the  general case

In the second, and more interesting case because
often found in telescope optics, the two mirrors share the plane of incidence. The Mueller matrix is then just the multiplication of the
two mirror Mueller matrices, $M_{M1}M_{M2}$, and it conserves the 
block-diagonal symmetry. Even better the new matrix can be written
as the matrix of a single mirror
\begin{equation}
 M_{mirror}=\frac{1}{2}\left( \begin{matrix}  1+\chi_T^2 & 1-\chi_T^2 & 0 & 0 \\ 1-\chi_T^2 & 1+\chi_T^2 & 0 & 0 \\ 0 & 0 & 2\chi_T\cos \Delta_T & 2\chi_T\sin \Delta_T  
\\ 0 & 0 & -2\chi_T\sin \Delta_T& 2\chi_T\cos \Delta_T \end{matrix} \right)
\end{equation}
where $\chi_T=\prod_{i=1,N} \chi_i$ and $\Delta_T=\sum_{i=1,N} \Delta_i$ for any number $N$ of mirrors sharing the plane of incidence. 
Our method can therefore be easily generalize to an optical train of mirrors sharing the same plane of incidence and for which we will determine
easily the compound $\chi_T$ and $\Delta_T$ of the equivalent mirror. It may prove impossible to determine the individual values of each single mirror, unless
they are identical mirrors. But this should not be a worry, for the goal was to calibrate the system and not to measure the mirrors individually.

Finally, there are many configurations in which we can make use of the following trick. Let us suppose that only two mirrors are in the 
optical train to be calibrated, and that we can modify
the relative angle $\theta$ between their planes of incidence. Let us know point at a source of 
unpolarized light. This may be the solar disk for solar telescopes, or any of the planetary targets listed in Section 3 
(Moon, Mercury, Mars, asteroids) at opposition (phase 0). The effect of the first
mirror on unpolarized light is to introduce linear polarization. We do not know  the amount of linear polarization
before calibrating that first mirror, but we know  its orientation with full certainty: it is the plane of
incidence of the mirror. Hence the
first mirror behaves for the second mirror as one
of the targets we listed in Sect.3 \citep{borra_polarimetry_1976}: a source of linear polarization with known orientation but unknown rate of 
polarization. Since we assumed
that we could change the angle $\theta$ at will, it plays now the role of angle $\alpha$ in our method and we can proceed to calibrate mirror 2. Once 
we are satisfied with this calibration we are left with mirror 1 alone, which can be calibrated by using the listed targets.

   
%
   
\section{Conclusion}

In an increasing number of modern telescopes the  access to polarization-free foci becomes almost impossible for polarimetry. At the same time
the larger and larger apertures of these instruments make imposible to put polarizers and retarders in front of the main aperture to calibrate them. 
The need appears for methods to calibrate these instruments that do not require 
modifications of the telescope optics. For night observations, lists of calibration targets have been compiled with
suposedly known polarizations that can be used as sources. But we appreciate
two critical problems with them: they are not accessible in dayside for solar observations, and the calibration depends on trusting that the polarization
from that source is the one listed and not other.

The physics of polarization of light shows that the degree of polarization depends 
on many different details impossible to control on astronomical objects. On 
the other hand there are a number of situations on which the orientation of 
the linear polarization can be determined out of geometrical arguments alone.

In this work we propose a method that uses that sole information from the 
observed target: there is linear polarization at an angle which can be
perfectly determined and that we can change at will. In our favourite case, the polarization of the continuum spectrum reflected off the Moon
will be perpendicular to the plane Sun-Moon-Earth, while its amplitude can vary depending of a large amount of factors. By rotating the
telescope (or the mirror to be calibrated) so that this linear polarization incides in our mirror at all possible angles respect to the 
incident plane we can compile
enough information to calibrate a mirror. The nice symmetries of the Mueller matrix of a mirror help making this possible.

Calibrating a mirror without the requirement of knowing the degree of polarization or without the use of circular polarization is a nice feat, but
it does not come for free. The measurements have to be done with  large accuracy and,
since the sources of astronomical polarization are far from being fully polarized, this implies large signal-to-noise ratios. We have illustrated
this with characteristic polarization levels of 1\% and 10\%, and see that
noise levels of $10^{-3}$ and $10^{-4}$ times the intensity are the limit for 
a sound polarimetric calibration.

Yet we anticipate that this technique is easy and robust enough to implement that it may be of interest for many ground and space
telescopes, including
those cases of multiple mirrors as long as 
they are in the same plane.
\newpage

\begin{acknowledgements}
I gratefully acknowledge the help of F. Paletou (IRAP) in reading and correcting many of the errors in this work.
      This work was carried out in the framework of EU COST network MP1104 "Polarisation as a tool to
study the Solar System and beyond"; www.polarisation.eu.
\end{acknowledgements}

\bibliographystyle{aa}




\newpage



\section*{Supplementary material}

# AA201525991 Referee report 2

## March 25, 2015

This second evaluation report is in two parts. The first part addresses a persistent issue revolving about what is measured, what is a contrast, and how the desired parameters can be extracted. The second part deals with other aspects that were left aside in the first report, since the highest priority was placed on serious issues.

All mentions of text location in this report, like "below figure X", refer to the referee layout of resubmitted version 2.

# 1 From polarimetric measurements to the mirror's Mueller matrix

## 1.1 No photometric calibration

The author makes this point, with which the referee fully agrees.

*Actually at this point we are going to impose a further requirement and it is that we do not depend on I or the actual intensity measurement in our detectors. We impose these restrictions because there is no reason to pretend that we know the exact intensity coming from our celestial object and we cannot let the calibration depend on the accurate photometric calibration of our instruments.*

## 1.2 What are Stokes parameters?

Each of the Stokes parameters, usually with notation $I, Q, U, V$ is in *energetic* units, either power (Watt, etc) or intensity (Watt/m2/sr, etc) with more variants for spectral densities. The author uses the same notations for the available measurements (Eqn. 5-8 in submitted v2), and the referee believes that what on the surface looks like a trivial notation issue is one of the root causes of the problem.

### 1.3   What are actually the measurements?

The author comments about the possible configurations for the polarimeter.

*Often the polarization analysis is made by a polarizing beamsplitter as a Savart plate, so that the detector receives two beams with intensities $I \pm S$...*

Note that if two beams are separated, there are two detector**s** involved. There may even be four, following:

*And in the most performant polarimeters all this is done simultaneously*

Each of these has its own gain $g$, or calibration, whether volt/Watt, or ADU/Watt (ADU: analog-digital converter units). Consistent with the statement in 1.1 above, the measurements consist in:

$$V_{out1}(\alpha) = g_1 \times (I + Q\cos(\alpha))$$

and similar equations for the other three channels. The method to derive the parameters $\chi$ and $\Delta$ must include an elimination of the irrelevant unknowns $g_1 \cdots g_4$.

### 1.4   A contrast is a ratio, not a difference

The concern to define optical quantities independent of power calibration is of course not new, and underlies the definition of fringe contrast (more commonly called fringe visibility in an interferometric context, but the idea is the same). So the fringe contrast (/visibility) is defined as a *ratio*. See, e.g.:

Equation (7.8) may then be rewritten as

$$I(\xi) = I_0\big(1 + \mathrm{Re}\{\gamma(x_1, y_1; x_2, y_2)\exp[\mathrm{i}2\pi d\xi/(\lambda z_0)]\}\big). \qquad (7.10)$$

A comparison of Eqs. (7.10) and (7.6) reveals that the fringe contrast produced by the pinholes at $(x_1, y_1)$ and $(x_2, y_2)$ is equal to $|\gamma|$ and that the phase of $\gamma$ determines the shift of these fringes from the center. The function $\gamma$ is thus described as the complex degree of spatial coherence between $(x_1, y_1)$ and $(x_2, y_2)$. Substituting expression (7.7) for $\Delta\phi$ in Eq. (7.9c) yields

$$\gamma(x_1, y_1; x_2, y_2) = \exp\big\{\mathrm{i}\pi[(x_1^2 + y_1^2) - (x_2^2 + y_2^2)]/(\lambda z_s)\big\}$$
$$\times \int_{\text{source}} \tilde{I}(x, y)\exp\{-\mathrm{i}2\pi[(x_1 - x_2)x$$
$$+ (y_1 - y_2)y]/(\lambda z_s)\}\,\mathrm{d}x\,\mathrm{d}y. \quad (7.11)$$

https://books.google.fr/books?id=Nrkfmf4RizIC
http://scienceworld.wolfram.com/physics/FringeVisibility.html
http://en.wikipedia.org/wiki/Interferometric_visibility

that show that visibility (contrast) is obtained as the ratio of the modulation amplitude by the mean intensity.

For completeness, it must be mentioned that the subtraction of a sliding mean can be used, e.g. in the case of a summing interferometer, to isolate the pure interference pattern.

## 1.5 Subtracting the mean does not help

Author (middle page 6, referee layout) states:
*This constant level can however be easily subtracted from both measurements independently by averaging over all the measurements. And, once this average is subtracted, the amplitude of the variation, the contrast, is independent of the photometric calibration.*

Subtracting the mean value from Eqn. 5 and 6 gives the respective quantities $g_1 Q \cos(\alpha)$ and $-g_2 \chi^2 Q \cos(\alpha)$ from which it is not possible to derive either $Q$ or $\chi^2$. At some point the author seems to have realized that taking a ratio might be useful: the ordinate of Figure 1 is
*Signal modulation over average*
which is disconcerting, since the vertical axis scale is centered around zero, more consistent with the *subtraction* mentioned in the text.

Another source of confusion is when the author uses the symbol $Q$ (a Stokes parameter, power or intensity) associated with the name *degree of polarization*, in the paragraph below Eqn 5-8, and also in the third paragraph of Sect. 3 (authors's responsibility to check whole article). A more proper definition of the degree of polarization, for a source known to be linearly polarized ($V = 0$) and with the reference frame aligned with the source's polarization axes ($U = 0$) would be $q = Q/I$, independent of telescope aperture, detector efficiency, etc... And, in fact, this is implied by the appearance in the corner of Fig. 4 and Fig. 5 of the fraction $Q/I$ presumably to quantify the degree of polarization. The referee confesses having been confused himself (by the text), writing in the first report, that the first equation could provide $Q$, should have written $q = Q/I$.

## 1.6 Yet a solution is possible

As stated above, Eqns (5-6) do not allow to determine $\chi$. But all is not lost. The author's remarks (below Eqn 12)
*This phase shift has different sign in one and other measurements $(I \pm U)_{out}$ providing extra redundancy.*
might have alerted him that there was some unused information. Below is a synopsis for a truly calibration-independent derivation of $Q$ and $\chi^2$.

1. As above, let $q = Q/I$.

2. Normalize Eqn 5 by mean value, resulting: $y_5(\alpha) = 1 + q\cos(alpha)$, straightforward data reduction provides the value of $q$.

3. Similar procedure with Eqn 6 give no extra information (at least in a noise-free context).

4. Divide Eqns (7-8) by their constant terms, resulting in:
   $y_7 = 1 + q \times \left( \frac{1-\chi^2}{1+\chi^2} \cos\alpha + 2\frac{\chi}{1+\chi^2} \cos\Delta \sin\alpha \right)$;
   $y_8 = 1 + q \times \left( \frac{1-\chi^2}{1+\chi^2} \cos\alpha - 2\frac{\chi}{1+\chi^2} \cos\Delta \sin\alpha \right)$;

5. Subtract constant part (the leading "1")

6. divide by $q$ (known from previous steps)

7. take half-sum and half-difference, to isolate: $z_1 = \frac{1-\chi^2}{1+\chi^2} \cos\alpha$
   and $z_2 = 2\frac{\chi}{1+\chi^2} \cos\Delta \sin\alpha$

8. cross-check: that the two latter expressions indeed have the proper dependence in $\alpha$ (phase, absence from other harmonics, etc)

9. Fit a $\cos\alpha$ dependence to the $z_1$ curve, use amplitude to solve for $\chi^2$. Note: sign of $\chi$ is ambiguous.

10. Fit a $\sin\alpha$ dependence to the $z_2$ curve, obtain amplitude, substitute value of $\chi$ and solve for $\cos\Delta$

Note that the derived values have the following ambiguities:

$$\Delta \longleftrightarrow -\Delta$$

and

$$\left. \begin{array}{c} \chi \\ \Delta \end{array} \right\} \longleftrightarrow \left\{ \begin{array}{c} -\chi \\ \Delta + \pi \end{array} \right.$$

The first one is inherent in the measurement scheme, where one essentially measures the axial ratio of an ellipse without knowing if it is right- or left-handed. The second one boils down to a choice of sign conventions for the *in* and *out* frames in a mirror reflection; in all reasonable choices, the sign of $\chi$ will be physically required to be positive. Also note that these ambiguities (symmetries) are not a deficiency of the solution proposed schematically above, since these symmetries are also obeyed by the set of equations (5-8).

## 1.7 Last remarks

Strictly speaking, the author claims for his method the independence from *absolute* photometric calibration, while this referee notes that the method fails in the absence of *relative* calibration of the various channels of the polarimeter. There is *in principle* room for a TBD polarimeter that would have excellent (tbd%) cross-calibration beween channels, allowing to derive $\chi^2$ more or less as proposed by the author. However:

- the author discusses the implementation of the polarimeter in only general terms and does not raise the issue of photometric cross-calibration of the channels (if the polarimeter has more than one physical detector);

- the method proposed above by this referee has no such requirements of cross-polarization;

- this referee's experience, even with a polarimeter that avoided the cross-calibration issue by using a single detector (and rotating crystalline plates) is a reproducibility of Stokes parameters of order of 1% on timescales of minutes in a temperature-controlled underground laboratory.

# 2 Other issues

## 2.1 Required accuracy

Several issues here.

**1. From measurement noise to parameter error**. Author states (below Fig.2, boldface): *In a Montecarlo test, Gaussian random noise of the amplitudes shown in the figure is added to the signals then they are fit with a Levenberg-Marquardt inversion algorithm. The quantity plotted is the average residual of $\chi^2$ along the Montecarlo series after the measure of the ratio of contrasts.*

First, language clarification: "average residual of $\chi^2$" should be "root-mean square difference between the derived and the true underlying value of $\chi^2$"; the word "residual" normally refers to something different: the differences between the model (for a trial value of $\chi^2$) and the measurements, along a data series similar to Fig.1. This discrepancy causes the reader to pause for comprehension.

Second, and more substantial: The "noise level" does not uniquely specify the conditions of a numerical simulation of parameter extraction as proposed by the author. For the same amount of noise on individual data points, the outcome of the numerical experiment will give a better accuracy on the estimate of $\chi^2$ for a larger number of measurements along the modulation curve; provided the curve is properly sampled, the rms error on $\chi^2$ scales as the inverse square root of the number of points.

**2. Correlated errors**. The argument above refers to errors on individual data points that are fully de-correlated. But there are also fully correlated errors (systematics). And all the in-between cases: error terms with some finite correlation time.

## 2.2   Available sources

The method, and the error simulation, assume that the dependence on alpha (twice the parallactic angle) can be sampled over a range of $2\pi$, allowing measurements similar to Fig.1. This will occur only for objects that transit near zenith at the telescope's geographical location. Less than optimum sampling still allows parameter extraction, but, with equal measurement accuracy, the rms accuracy of derived parameters will be worse for incompletely sampled curves. Even the incomplete sampling will take longer than for a transit near zenith, leaving more opportunities for various error-causing instrumental drifts. This restricts the availability of suitable sources for night telescopes.

For solar telescopes, the author states:
*the fact that to rerpoduce the rotation of the angle of polarization $\alpha$ it is sufficient to point at the limb at different position angles* note: typo, and $\alpha$ called the polarization angle, inconsistent with the author's special half/double convention
The method assumes that $Q$, or rather $Q/I$ is constant through the measurement. What are the expected variations of the degree of polarization $Q/I$ when going around the solar limb? And what is the impact on the extraction of mirror properties?

## 2.3   Special notation

The author uses a personal notation such that $\alpha$ appears where one expects $2\alpha$ in rotated Stokes vectors, Mueller matrices, etc. On the plus side the author's claimed advantage of simplifying equation by the elimination of a factor 2. On the negative side: readers confused by familiar equations that "do not look right". And the author himself fell into his own trap, when he states (top of page 5, referee layout):
*If we define $\alpha$ as half the known angle between the scattering plane and the incidence plane of the mirror*
when actually, if following his own logic, he should have written: ...***twice*** the known angle...



\end{document}